\documentclass[twocolumn,showpacs,preprintnumbers,amsmath,amssymb]{revtex4}

\usepackage{graphicx}
\usepackage{dcolumn}
\usepackage{bm}
\usepackage{epsfig}
\usepackage{float}

\begin{document}

\title{Molecular dynamics simulations of complex shaped particles using Minkowski operators.}

\author{S. A. Galindo-Torres }
\email{s.galinotorres@uq.edu.au}
\altaffiliation[Also at ]{Physics Department, Grupo de Simulacion de Sistemas Fisicos. Universidad Nacional de Colombia.}
\author{F. Alonso-Marroqu\'{\i}n}
\email{fernando@esscc.uq.edu.au}
\affiliation{ ESSCC \& School of Physical Sciences, \\
               The University of Queensland, Qld. 4068, Brisbane, Australia}

\date{\today}

\begin{abstract}
The Minkowski operators (addition and substraction of sets in vectorial spaces) has been extensively
used for Computer Graphics and Image Processing to represent complex shapes. Here we propose to
apply those mathematical concepts to extend the Molecular Dynamics (MD) Methods for simulations
with complex-shaped particles. A new concept of Voronoi-Minkowski diagrams is introduced to generate
random packings of complex-shaped particles with tunable particle roundness. By extending the classical
concept of Verlet list we achieve numerical efficiencies that do not grow quadratically with the body number of sides. Simulations of dissipative granular materials under shear demonstrate that the
method complies with the first law of thermodynamics for energy balance.
\end{abstract}


\maketitle

\section{Introduction}
It has passed more that $50$ years since Loup Verlet published the milestone paper which gives birth
to the Molecular Dynamics (MD) method \cite{verlet1967cec}. MD is the paradigm of modelling complex systems
as a  collection of particles interacting each other.  The method has been successfully applied for
computer simulations in different areas of Physics, Chemistry and Biology. Even engineers have also
recognized the potential of this method, after the pioneer work of Peter Cundall in the extension of
MD to model dissipative particle materials. Cundall's ideas on the use of discrete approach for
modeling rocks and soils were proposed in the appendix of his PhD thesis in $1971$ \cite{cundall71}. This
appendix led to a new area of numerical analysis of engineering problems, which is today known as
Discrete Element Method (DEM).  Most of the advances in the areas of MD and DEM have been based on the
development  of efficient and robust methods to account particle interaction. The advances in modeling
of the real particle shape, however, has been rather slower, in part because the existing methods require
too much computational effort, and partly because the difficulty to achieve physical correctness in the
interactions. Indeed most of the current models use agglomerates of spheres to represent particles with
complex shape \cite{lu07,ryckaert1977nic}.

On the other hand, There is another fast growing area of research in computer graphics and rendering.
New developments are fuelled by computer games industry and special effects companies.
There is therefore  a natural interest to put these development to a good use of solving scientific
problems. The aim of this paper is to merge the molecular dynamics approach with some key elements
of computer graphics:  The Minkowski operators, Voronoi diagrams and Euclidian distance formulas.
This combination gives rise to a new molecular dynamics method that accounts both particle shape and
interactions, and at the same time keep a reasonable  balance between accuracy and efficiency.

Contrary from the computer games, scientific and engineering applications of numerical simulations
require high precision and interaction models which comply with the conservation principles of physics.
Simulations with large number of particles has been implemented using  three different methodologies:
The first one  corresponds to the so-called event-driven methods, where the interaction of the particles is
handling via collisions \cite{alam2003rbg}. These methods are suitable for loose particle materials, but it cannot
be applied for dense systems because they cannot handle permanent or lasting contacts \cite{mcnamara1996dfe}.
The second methodology is the Contact Dynamic methods \cite{jean1999nsc}.
In this method the equilibrium equations of the
system is solved in each time step. The method is suitable for resting contacts with infinite
stiffness, but simulations  are computationally expensive and lead in some cases to
indeterminacy in the solution  of contact forces \cite{mcnamara2004mip}. This
indeterminacy is removed by the method of Soft Molecular Dynamics\cite{mcnamara2005iao}
. In this method the  bodies are allowed to interpenetrate each other and the force is calculated in terms
of their overlap. The method has been successfully applied for circular and spherical
particles. However,  the determination of contact force for non-spherical particles is
still heuristic, computational inefficient, and in some cases it lacks physical correctness
\cite{hasegawa04,poeschel04}.

Soft Molecular Dynamics Methods has been proposed for two-dimensional (2D) simulations
using polygons \cite{alonso04c,poeschel04}. The interaction between the polygons is
calculated in terms of  the overlap area.  The  main drawback of this approach is that it
does not provides energy balance equations, as the elastic force between the polygons
does not belong from a potential \cite{poeschel04}. The approach of deriving the interaction
in terms of the overlap area is are also extremely difficult  to extend to 3D, because the
calculation of  the overlap  between two polyhedra is computationally  very expensive. A
solution of this difficulties  has been proposed recently using spheropolygons \cite{alonso08epl}.
These are figures that are obtained using the mathematical concept of Minkowski sum, first
introduced by Liebling and Pourning in the modeling of granular materials with non-spherical
particles \cite{pournin05b}.

This paper presents an extension of the original ideas of Verlet, Cundall, Pournin and
Liebling to model particulate materials with complex particle shape. In Section \ref{comgeom}
we begin introducing kew elements of Computer Graphics. Then we introduce the  new method of
Voronoi-Minkowski diagrams to generate packings of complex-shaped particles.  Section
\ref{verlet} extends the concept of Verlet list for neighbor identification between the
particles. In section \ref{sim} we perform numerical simulations of sheared
granular materials with elastic, frictional contacts. We deal with energy balance and numerical
efficiency.  In Section \ref{conclusions} we  present the perspectives of this work to extend the model
to 3D simulations.

\section{computational Geometry}
\label{comgeom}

Here we introduce some mathematical concepts which are useful to represent particle
shape.  After define the Minkowski operators and the Voronoi diagrams,  we apply these
two concepts to generate random packings of non-spherical particles.

\subsection{Minkowski operators}

Minkowski addition and substraction of sets belong to the area of Mathematical morphology.
This is a tool for image processing originally developed for quantitative description of
geological data \cite{serra1983iaa}. The primary application of morphology occurs in by applying
expanding and shrinking operations on binary or gray level images. Here we present the
definition of these operations in Euclidean spaces and it application to the generation
of objects with complex shape with tunable particle roundness.

\subsubsection{Dilation}
Given  two sets of points A and B in an Euclidean space, their Minkowski sum, or dilation,
is given  by

\begin{equation}
  A \oplus B=\{\vec x+ \vec y~|~ \vec x \in A,~ \vec y \in B \}
\end{equation}

\begin{figure}[h]
  \begin{center}
    \epsfig{file= 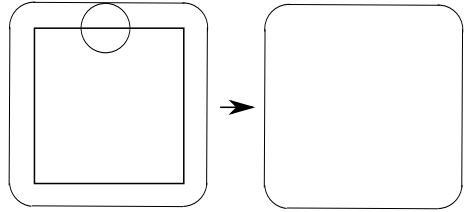,width=0.7\linewidth}
   \caption{Dilation of a square by a disk.}
   \label{fig:dilation}
  \end{center}
\end{figure}

This operation is geometrically equivalent to the sweeping of one set around
the profile of the other without changing the relative orientation. This paper
will deal with the spheropolygons, which are Minkowski sum of a polygon with a
disk, see Fig. \ref{fig:dilation}.
Other examples of  Minkowski sum operations are  the spherocyllinder
(sphere $\oplus$ line segment) \cite{pournin05a},  the Minkowski cow
(non-convex polygon $\oplus$ disk) \cite{alonso04b}, the spherosimplex
(sphere $\oplus$ simplex)  \cite{pournin05b} and the spheropolyhedron
(sphere $\oplus$ polyhedron)  \cite{pournin05c}.

\subsubsection{Erosion}
The erosion of a set A by the set B is defined over an Euclidean space E by:

\begin{equation}
A \ominus B = \{\vec x | B_{\vec x} \subseteq A\},
\end{equation}

where $B_{\vec x}$ is the translation of B by the vector $\vec x$, i.e.,

\begin{equation}
B_{\vec x} = \{\vec y + \vec x | \vec y \in B\}.
\end{equation}

\begin{figure}[h]
  \begin{center}
    \epsfig{file= 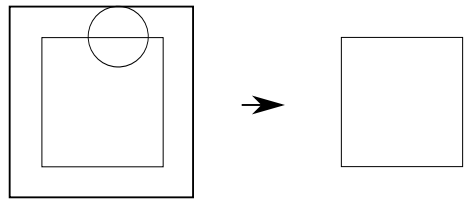,width=0.7\linewidth}
   \caption{Erosion of a square by a disk}
   \label{fig:erosion}
  \end{center}
\end{figure}

Then the erosion of A by B can be understood as the locus of points reached by the center of B
when B moves inside A. If A is a polygon and B is a disk of radius $r$, the erosion is a polygon
inside A  whose borders lie a distance $r$ from A, see Fig. \ref{fig:erosion}.

\subsubsection{Opening}
The opening of A by B is obtained by the erosion of A by B, followed by dilation of the resulting image by
B:

\begin{equation}
A \circ B  = (A \ominus B) \oplus B.
\end{equation}

\begin{figure}[h]
  \begin{center}
    \epsfig{file= 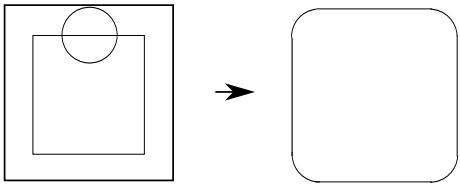,width=0.7\linewidth}
   \caption{Opening of a square by a disk.}
   \label{fig:opening}
  \end{center}
\end{figure}

In the case of a polygon, the opening by a disk is the polygon with rounded  corners, see Fig. \ref{fig:opening}. The degree of  roundness increases as the radius of the disk increases.

\subsection{Calculation of the mass properties of spheropolygons.}
\begin{figure}[h]
  \begin{center}
    \epsfig{file= 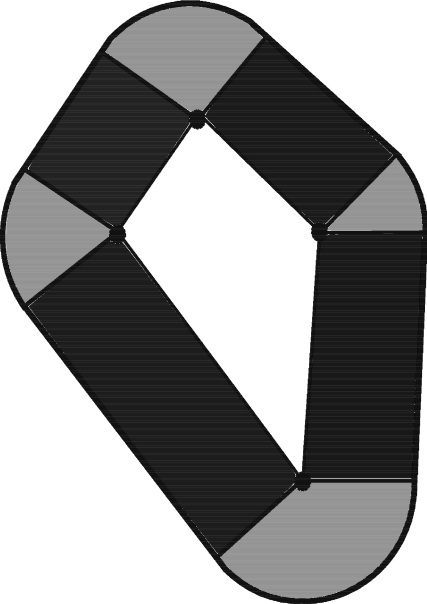,width=0.3\linewidth}
   \caption{General spheropolygon divided in special regions for the calculation of its mass properties.}
   \label{fig:massproperties}
  \end{center}
\end{figure}

Here we introduce an analytical  method to calculate the mass, center of mass and moment of inertia of the spheropolygons. Consider the general case of Fig. \ref{fig:massproperties}, the general spheropolygon is divided in some constituting parts.

The division is as follows, first in the center (white region in Fig. \ref{fig:massproperties}) we have the original polygon that will be dilated by the disk element. Then we have in black some rectangular sections of thickness equal to the dilation radius. Finally the disk sectors in gray have an angle equal to $\beta=\theta-\pi$ with $\theta$ the internal angle between the corresponding two segments of the original polygon.

The spheropolygon area, and therefore its mass, is easily calculated as the addition of the area of each sector. The area of rectangles is simply $A_{r}=l \times R$ where $l$ is the length of the side and $R$ the spheroradius. The area of the circular  sectors is $A_{c}=\frac{\beta}{2} R^2$. And the area of the central polygon is given by
\begin{equation}
A_{p}=\frac{1}{2}\sum_{i=0}^{n-1} (x_{i}y_{i+1}-x_{i+1}y_{i}),
\end{equation}

with $n$ the number of sides and $\vec{r}_{i}=(x_i,y_i)$ the coordinates of the i-th vertex.

To calculate the center of mass of the spheropolygon we start calculating the center of mass of each sector. The rectangular sectors are the easiest for this task, the center of mass is in the middle of them. The circular sectors have a center of mass in the bisector line a distance $d_{cm}=\frac{4 R \sin(\beta/2)}{3\beta}$ from the center of the circle. The coordinates for the central polygon center of mass are
\begin{eqnarray}
x_{cm}=\frac{1}{6 A_{p}}\sum_{i=0}^{n-1}(x_i+x_{i+1})(x_{i}y_{i+1}-x_{i+1}y_{i}) \\
y_{cm}=\frac{1}{6 A_{p}}\sum_{i=0}^{n-1}(y_i+y_{i+1})(x_{i}y_{i+1}-x_{i+1}y_{i}).
\end{eqnarray}
The total center of mass is the weighted sum of all these center of mass coordinates.

Now we calculate the moment of inertia. Since this is a 2D problem, the moment of inertia is just a scalar quantity. First we calculate the moment of inertia around the center of mass of all the regions. For the rectangles it is known that $I_{r}=\frac{1}{12}M_{r}(l^2+R^2)$ with $M_{r}$ the rectangle mass. The moment of inertia for the circular sector is
\begin{equation}
I_{c}=\frac{M_{c}R^2(16(\cos(\beta)-1)+9\beta^2)}{18\beta^2}.
\end{equation}
Finally for the polygon moment of inertia the formula is
\begin{equation}
I_{p}=\frac{M_{p}\sum_{i=0}^{n-1} \| \vec{r}_{i} \times \vec{r}_{i+1}\|(r_{i}^2+\vec{r}_{i} \cdot \vec{r}_{i+1}+r_{i+1}^2)}{6\sum_{i=0}^{n-1}\| \vec{r}_{i} \times \vec{r}_{i+1}\|}.
\end{equation}

The total moment of inertia is given by the addition of all these different quantities with the parallel axis theorem. This theorem states that the moment of inertia of a body with respect to an axis separated a distance $L$ from its center of mass is
\begin{equation}
I=I_{cm}+ML^2,
\end{equation}
where $I_{cm}$ is the moment of inertia of the body referred to its center of mass and $M$ is its mass. With this the mass properties calculations are complete.
\subsection{Voronoi-Minkowski diagrams}

Here we detail the method to generate random packings of spheropolygons  with  a tuning void ratio and
particle roundness.  This approach uses the concept of Voronoi diagrams. This is a special decomposition of the Euclidean space by space filling polygons.  These polygons are generated by choosing a set of points in the space, which are called the Voronoi  sites. Each  site $\vec p$ has a Voronoi cell, which consists of all points closer to $\vec p$ than to any other site.

The cells of the Voronoi construction are polygons in contact with each other without void spaces. Cells with void spaces and roundness will be constructed by using the concept of Voronoi-Minkowski diagrams. They are collections of spheropolygons resulting from the application of  Minkowski operations to each Voronoi Cell. The operations are chosen such that each spheropolygon lies entirely in the voronoi cells, so that the spheropolygons has no overlap.

\begin{figure}[h]
  \begin{center}
    \epsfig{file= 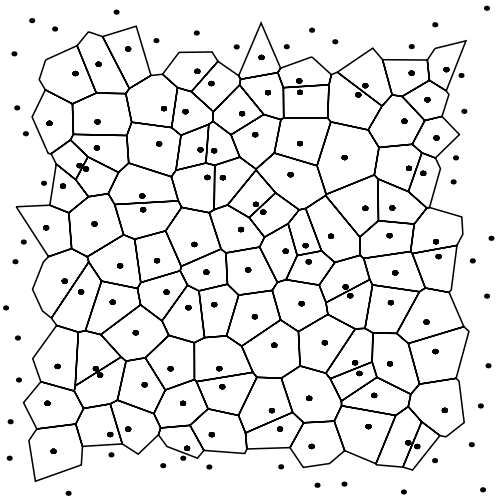,width=0.8\linewidth}
   \caption{ A  typical voronoi construction. The Voronoi sites are randomly distributed in the plane and the polygons are the set of closest point to each sites.}
   \label{fig:voronoi}
  \end{center}
\end{figure}

The implementation of the Voronoi-Minkowski cells in 2D consists of two steps. To begin we generate the voronoi cells. We set the Voronoi sites randomly distributed in a plane. Then the polygons are constructed by searching
the set of closest points to a particular Voronoi point than to any other. The resulting Voronoi cells has a random distribution of size and shape, as shown the  Fig. \ref{fig:voronoi}.

The second step generates an spheropolygon $W_i$ in each Voronoi cell $A_i$. We first erode the cell by a disk $B_{d}$ of radius $d$. Then we apply the morphological dilation on the eroded polygon by a disk $B_{r}$.

\begin{equation}
W_i  = (A_i \ominus B_{d}) \oplus B_{r}.
\label{eq:erosion-dilation}
\end{equation}

The condition $r \le d$ guarantees that the spheropolygon is a subset of the Voronoi cell. If $r = d$, the
Eq. (\ref{eq:erosion-dilation}) reduces to the opening defined above. In this case the spheropolygons touch with the neighbors
by their borders.

\begin{figure}[h]
  \begin{center}
    \epsfig{file= 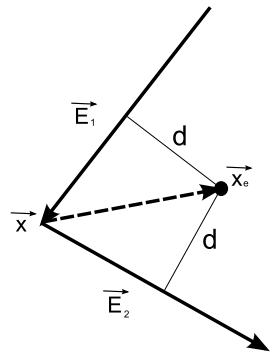,width=0.4\linewidth}

   \caption{The erosion of the voronoi polygon is done by finding the sub polygon vertices considering that they are at distance $d$ of two sides of the polygon. ${\bf E}_{1}$ and ${\bf E}_{2}$ are the vector representing two different edges of the voronoi polygon. The dashed vector goes from the intersection to the vertices of the sub polygon }
   \label{fig:erosion1}
  \end{center}
\end{figure}

The implementation of the erosion of each Voronoi cell is described as follows: For each vertex
of the cell we find a sub-polygon on the inside of the voronoi polygon. Each vertex $\vec x_e$ of this sub-polygon
satisfies the criterion that each one of them is exactly at a distance $d$ of two different sides of the voronoi
polygon. As can be seen in Fig. \ref{fig:erosion1} the point $\vec x_e$ is obtained as the vector:

\begin{equation}
\vec x_e = \vec x + d \cot(\theta/2) \hat{e}_2 +d (\hat k \times \hat{e}_2).
\label{eq:inside point}
\end{equation}

\begin{figure}[h]
  \begin{center}
    \epsfig{file= 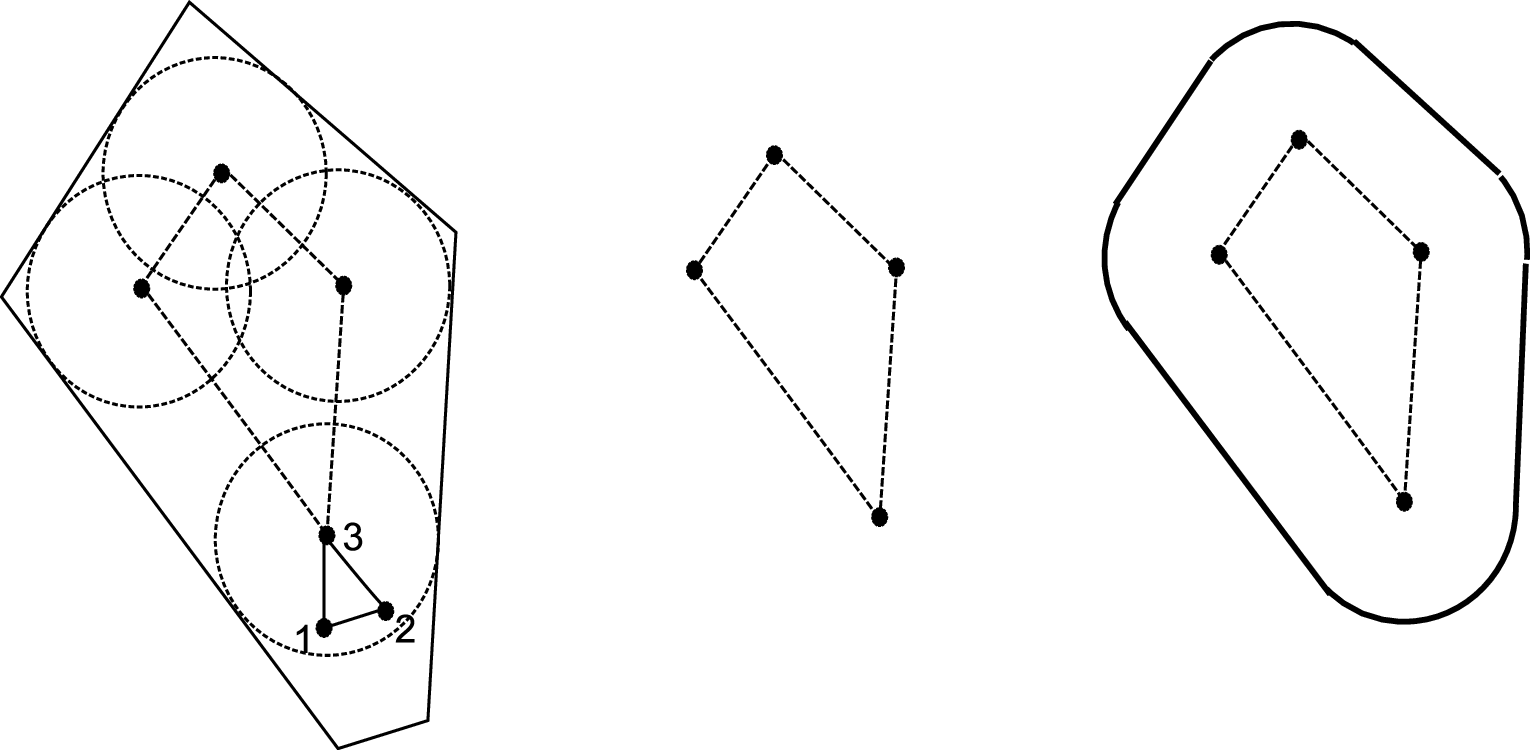,width=\linewidth}
   \caption{The initial voronoi polygon $A$ (left) is eroded by a disk element $B_d$ of radius $d$. The eroded polygon $A\ominus B_d$ (center) is then dilated by the same disk element $B_d$ producing the desired spheropolygon $(A \ominus B_d)\oplus B_d$ (right). Note the pathological case of points 1 and 2, since they are closer than $d$ to one of the polygon sides, they are substituted by the point 3.}
   \label{fig:opening Voronoi}
  \end{center}
\end{figure}

Where $\vec x$ is the position of the vertex of the polygon, $\hat{e}_2$ is the unitary vector of the edge $E_2$, $\theta$ is the angle between the edges $E_1$ and $E_2$ and  $\hat k$ is a vector perpendicular to the plane. Since the set of polygon vertices are arrange in a counterclockwise orientation the product
$\hat k \times \hat{e}_2$ points always to the inside of the polygon. As shown the Fig. \ref{fig:opening Voronoi}, the set of points satisfying Eq. (\ref{eq:inside point}) gives us a possible vertex of the erosion sub polygon. After we chose only those points which are not
closer than $d$ to any other polygon side, as shown in Fig. \ref{fig:opening Voronoi}.

\begin{figure}[h]
\begin{center}
  \epsfig{file=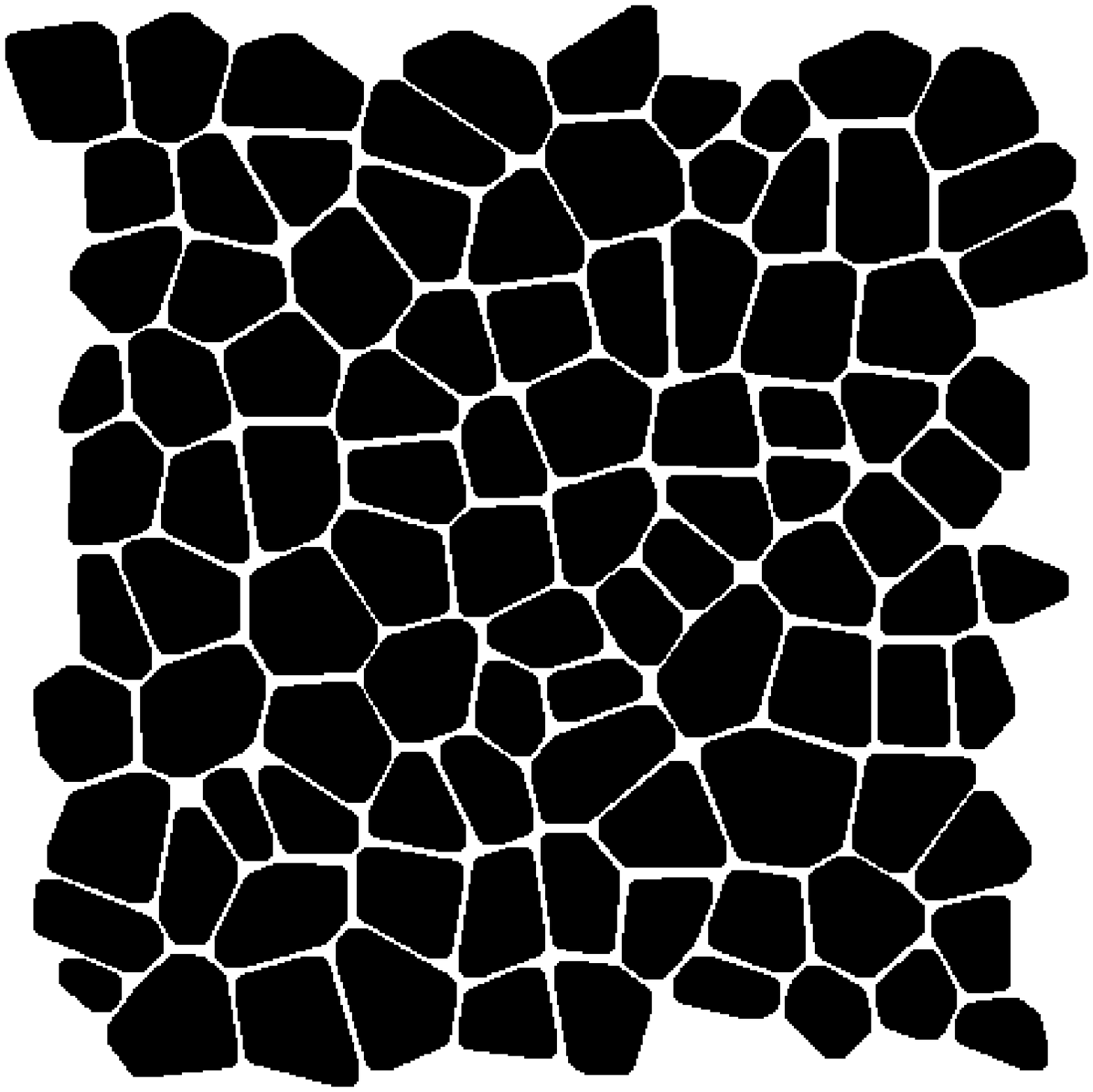,width=\linewidth}
\caption{An array of voronoi spheropolygons constructed by the method described above. The disk element used for the dilation process is slightly smaller than the one used in the erosion in order to allow void space between particles. In the opening the two disk elements are the same and there are no void spaces between sides shared by two polygons.}
   \label{fig:sphero}
\end{center}
\end{figure}

Once the operation given by Eq. (\ref{eq:erosion-dilation})  is done for all the voronoi cells, we can construct Voronoi-Minkowski diagrams like the one showed in Fig \ref{fig:sphero}. The parameter $d$ of Eq. (\ref{eq:erosion-dilation}) controls the initial packing of the system. The parameter $r$ is a measure of the angularity of the spheropolygons: as $r$ decreases the particles became more angular. Therefore the Voronoi-Minkowski diagrams allow us to generate random packings of particles with tunable void ratio and roundness.

\section{Interaction force}
\label{verlet}

To solve the interaction between spheropolygons we consider all vertex-edge distances between the polygons base.
Let's take two spheropolygons $W_i=P_i \oplus B_{r_i}$ and  $W_j=P_j \oplus B_{r_j}$. We will call $P_i$ and $P_j$ the polygons base, and $r_i$ and $r_j$ the sphero-radii. Each polygon is defined by the set of vertices $\{V_i\}$ and  edges $\{E_j\}$. The force $\vec F_{ij}$ acting on the spheropolygon $i$  by the spheropolygon $j$ is defined by:

\begin{equation}
\vec F_{ij}=-\vec F_{ji}=  \sum_{V_i E_j}{\vec F(V_i,E_j)}
               +\sum_{V_j E_i}{\vec F(V_j,E_i)},
\label{eq:contact force}
\end{equation}

\noindent

where $\vec F(V,E)$ is force between the vertex $V$ and the edge $E$. The torque
resulted from this force is given by

\begin{equation}
\begin{array}{clcr}
\tau_{ij} & = & \sum_{V_i E_j}{(\vec R(V_i,E_j)  - \vec r_i)
                        \times \vec F(E_i,V_j) } \\
          &+ &\sum_{V_j E_i}{ (\vec R(V_j,E_i) - \vec r_i)
                        \times \vec F(E_j,V_i) },
\end{array}
\label{eq:torque}
\end{equation}

\noindent
where $\vec r_i$ is the center of mass of the particle $i$ and $R(V,E)$ is the point of
application of the force and . This is the middle point of the overlap region between
the vertex and the edge:

\begin{equation}
\vec R(V,E) = \vec X + (r_i + \frac{1}{2} \delta(V,E))
\frac{\vec X-\vec Y}{||\vec Y-\vec X||},
\end{equation}

\noindent
where $\delta(V,E)$ is defined as

\begin{equation}
\delta(V,E)=\langle r_i+r_j-d(V,E) \rangle,
\label{eq:overlap}
\end{equation}

\noindent
and $d(X,E)= ||\vec Y-\vec X||$ is the distance from the vertex $V$ to the segment $E$.
Here  $\vec X$ is the position of the vertex $V$ and $\vec Y$ is its closest point
on the edge $E$. The ramp function  $\langle x\rangle$ returns $x$ if $x>0$
and zero otherwise.

The calculation of the force is extremely simple. It does not need neither calculations
of overlap areas between nor calculation of Minkowski operators. Further reduction of the
number of calculation is also possible by restricted the force calculation only on those
vertex-edge pairs  which are relative close each other.

With this aim we introduce the  {\it neighbor list}, which is the
collection of pair particles whose distance between them is less than
$2\alpha$.  The parameter $\alpha$ is equivalent to the {\it Verlet distance}
proposed by Verlet to speed up simulations with spherical particles~\cite{verlet1967cec}.

A {\it link cell} algotithm \cite{poeschel04} is  used to allow rapid calculation of this
neighbor  list. First we decompose the space domain of the simulation
into a collection of square cells of side $D+\alpha$, where $D$ is the
maximal diameter of  the particles. Each cell has a list of particles hosted in it.
Then the neighbors for each particle are searched only in the
cell occupied by this particle, and its eight neighbor cells.

For each element of the neighbor list we create {\it contact list}. This list consists of
those vertex-edge pairs whose distance between them is less than $r_i+r_j+2\alpha$,
where $r_i$ and $r_j$ are the sphero-radii.  In each time step, only these vertex-edge
pairs are involved in the contact  force calculations. Overall, the {\it Verlet list} for
neighborhood  identification consists of a neighbor list with all pair of neighbor
particles, and one contact list for each pair of neighbors. These lists require little
memory storage, and they reduce the amount of calculations of contact forces to $O(N)$,
which is of the same order as in simulations with spherical particles \cite{poeschel04}.

The {\it Verlet list} is calculated at the beginning of the simulation, and it is
updated when the following condition is satisfied:

\begin{equation}
 \max_{1 \le i \le N}\{\Delta x_i+R_i\Delta\theta_i\}>\alpha.
\label{eq:neighbor update condition}
\end{equation}

\noindent
$\Delta x_i$ and $\Delta\theta_i$ are the maximal displacement and rotation
of the particle after the last neighbor list update. $R_i$ the maximal distance
from the points on the particle to its center of mass.  After each update
$\Delta x_i$ and $\Delta\theta_i$ are set to zero. The update condition is
checked in each time step.

Increasing the value of  $\alpha$ makes  updating of the list less frequent,
but increases its size, and hence the number of force calculations. Therefore,
the parameter $\alpha$ must be chosen by chosing a reasonable balance between
the time expended by the force calculations and the updating of neighbor lists.

\section{Simulations}
\label{sim}

In order to check the validity of this method we created a shear band simulation for
our spheropolygons with periodic boundary conditions. The polygons are confined by
two horizontal plates. The lower plate is fixed. The upper plate has a fixed load and
a constant horizontal velocity. First he verify energy balance by calculating the
work done by the external forces, the dissipated energy, and the internal energy of
the systems. Then we perform a series of simulations to analize the computational
efficiency of the method.

\begin{figure}[h]
\begin{center}
\epsfig{file=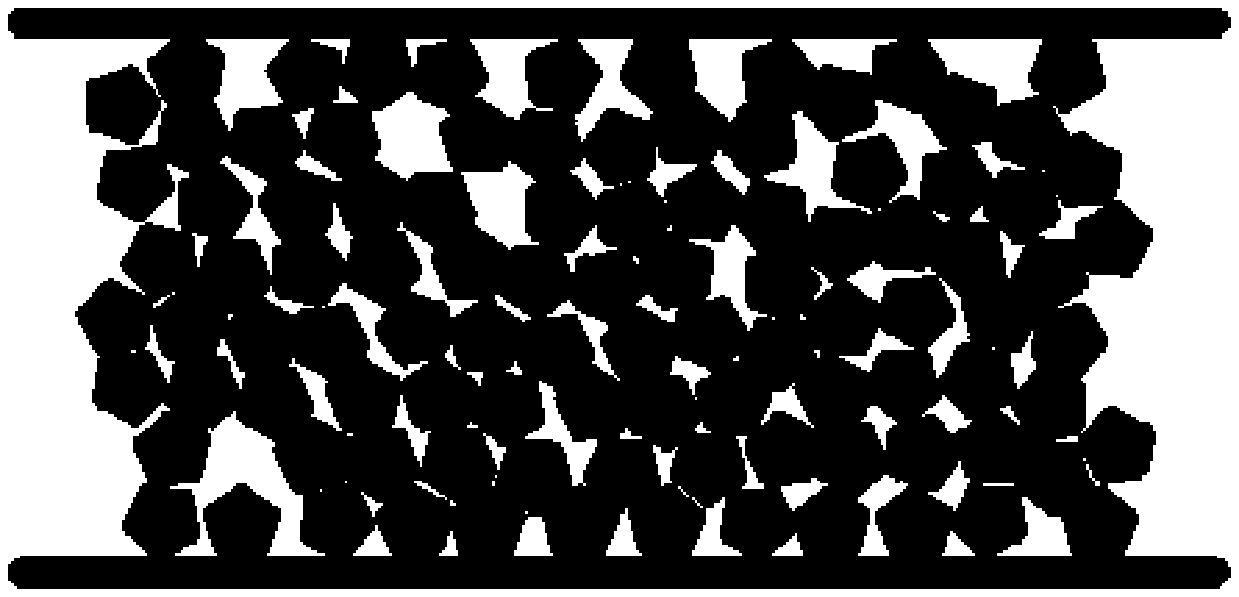,width=\linewidth}
\epsfig{file=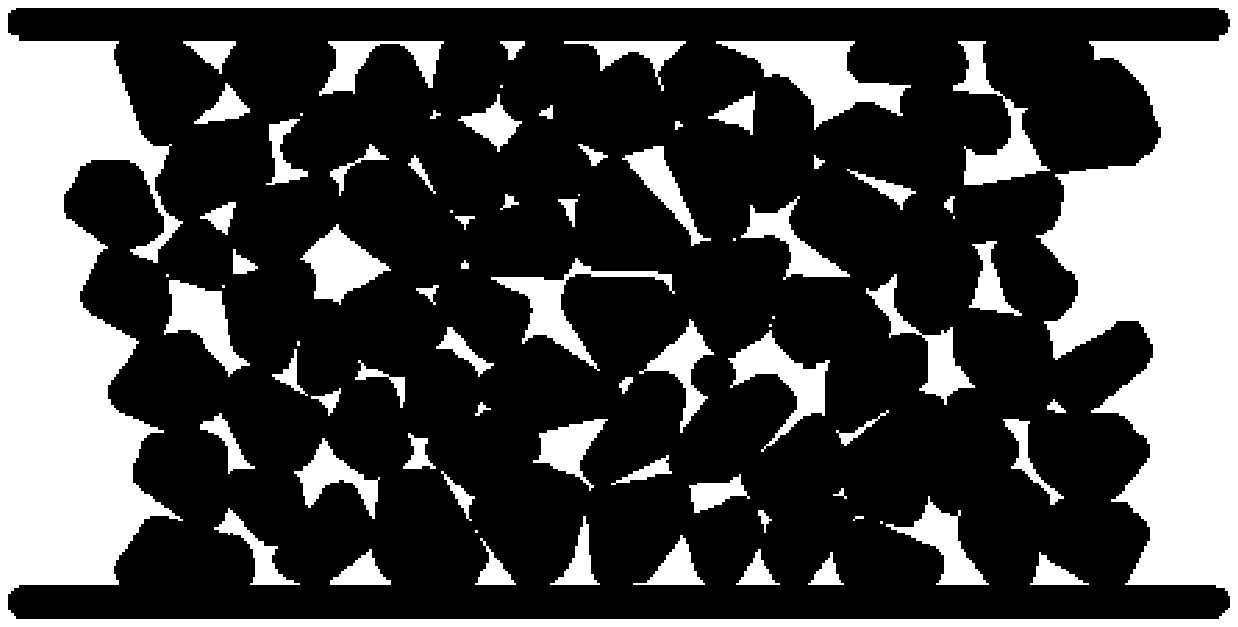,width=\linewidth}
\caption{Shear band simulator with regular pentagons (above) and voronoi polygons (below)}
\label{fig:shear}
\end{center}
\end{figure}

\subsection{Energy Balance Test}

In this section we present numerical results from simulations with dissipative granular
materials. We check whether the system complies with the energy balance as established by
the first law of the thermodynamics (work done by the system = energy dissipated
+ change of inernal energy of the system).

Simulations are performed using two different particle geometries: regular spheropolyhedra
and particles generated with Voronoi-Minkowski diagrams, see Fig. \ref{fig:shear} .
These systems are simulated by replacing in Eq.~\ref{eq:contact force} the following
repulsive, frictional force

\begin{equation}
\vec F(V,E)= k_n \delta_n \vec N + k_t \delta_t \vec T,
\end{equation}

\noindent
where $\vec N = \frac{\vec Y-\vec X}{||\vec Y-\vec X||}$ is the normal unit vector.
The tangential vector  $\vec T$ is taken perpendicular
to $\vec N$. The overlapping length $\delta_n$ is defined in
Eq.~\ref{eq:overlap}. The elastic displacement  $\delta_t $
accounts frictional forces, and it must satisfied the sliding
condition $|F_t|<\mu F_n$, where $\mu$ is the
coefficient of friction \cite{garciarojo2005cmr}.
The parameters of the simulations are the normal stiffness
$k=10000$, the tangential stiffness $k_t = 0.1k_n$, friction
coefficient $\mu =0.5$,  density $\sigma=1$,
and the time step $\Delta t = 0.00001$.
Viscosity forces proportional to relative velocity of the contacts
are also included to allows relaxation of the system.

The evolution of  $\vec r_i$ and the orientation  $\varphi_i$  of the particle
is obtained by solving the equations of motion:

\begin{equation}
 m_i\ddot{\vec{r}}_i  =\sum_{j}\vec F_{ji}-m_i g \hat y,  ~~~~~~
I_i\ddot{\varphi}_{i} =\sum_{j}{\tau_{ji}}.
\label{eq:newton}
\end{equation}

\noindent
Here $m_i$ and $I_i$ are the mass and moment of inertia of the particle.
The sum is over all particles interacting with this particle; $g=10$ is the
gravity; and $\hat y$ is the unit vector along the vertical direction.
The equations of motion of the system are numerically solved using a four order
predictor-corrector  algorithm \cite{gear1971niv}.

\begin{figure}[H]
\begin{center}
\includegraphics[bb = 0 0 1140 1614,scale=0.4,trim=0cm 0cm 15cm 35cm,clip]{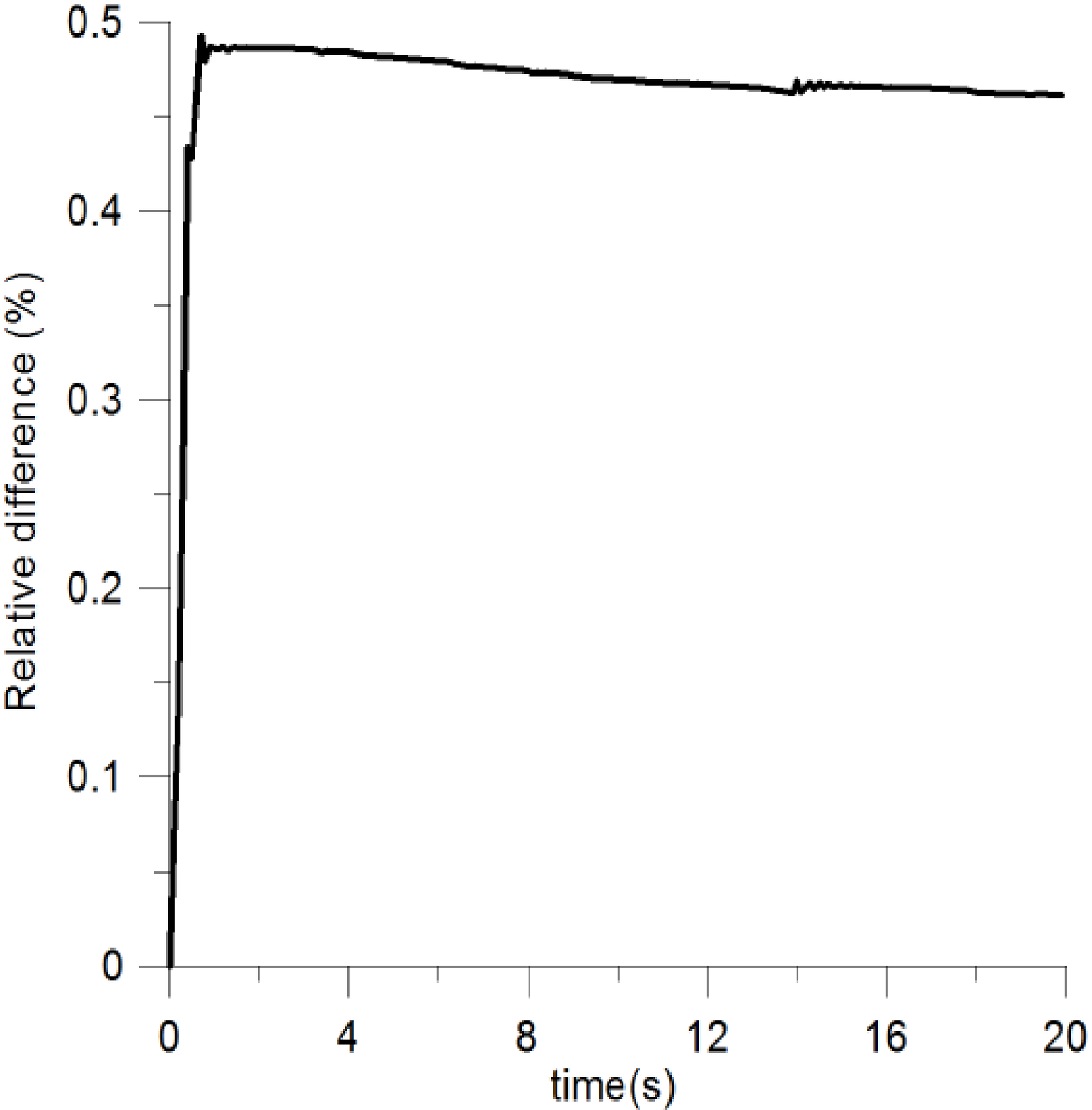}
\includegraphics[bb = 0 0 1140 1614,scale=0.4,trim=0cm 0cm 15cm 35cm,clip]{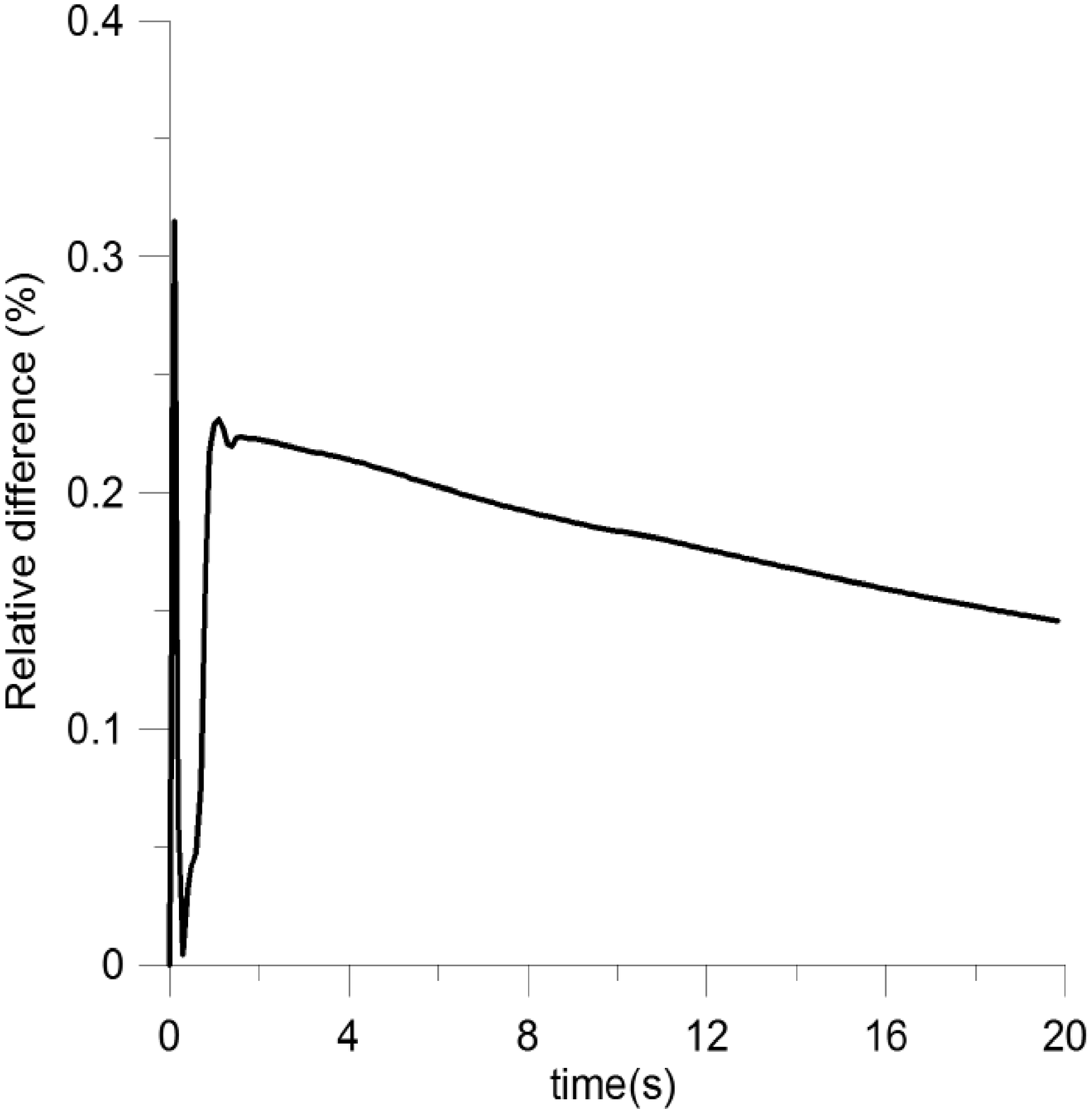}
\caption{Relative difference between the total mechanical energy and the dissipated one with the work by external forces for regular pentagons (above) and voronoi polygons (below)}
\label{fig:ediff}
\end{center}
\end{figure}

For both cases the total mechanical energy is obtained as well as the work done by the external forces. We show the relative difference for both cases (i.e. the difference between the mechanical energy plus de dissipated energy and the work done by external forces relative to the external work) in Fig. \ref{fig:ediff}.

As can be seen the difference is less than 1\% for both regular pentagons and voronoi polygons. The simulation is ready for problems that require energy measurement like the heating of internal water in the void spaces of the shear band or the increase of the rock temperature due to the frictional force. This is an important advantage of our model since the repulsive force as calculated provides a simple potential energy expression which was not
available in previous models \cite{alonso04c,poeschel04}.

\subsection{Performance Test}

One of the main optimization methods implemented in our code is the Verlet list. Certainly this method greatly reduce the amount of CPU time expended on the force calculations. The efficiency of the algorithm is meassure
using the so-called {\it Cundall number}\cite{alonso08b}. This number is the amount of particle time
steps executed by the processor in one second, which is calculated as $c = N_T N /T_{CPU}$, where $N_T$ is the number of time steps, $N$ is the number of particles and $T_{CPU}$ is the CPU time of the simulation.The
inverse of the Cundall number $1/c$ will be called here {\it normalized CPU time}  Previous simulations
\cite{alonso08b} shown that the Cundall number, although dependent on the processor, does not depend much on
the number of particle if $N>100$. Therefore for the simulation presented in this paper we chose samples
with $102$ particles.

\begin{figure}[H]
\begin{center}
\includegraphics[bb = 0 0 1140 1614,scale=0.4,trim=0cm 0cm 15cm 35cm,clip]{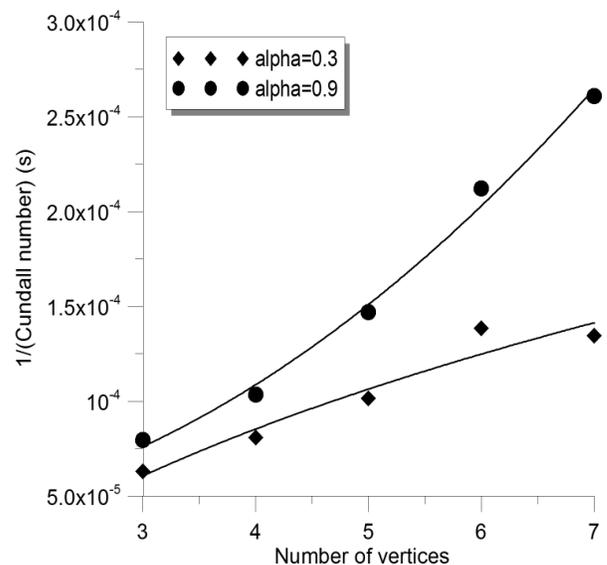}
\caption{Inverse of the Cundall Number versus number of vertices of the constituting element.}
\label{fig:geogranu}
\end{center}
\end{figure}

Our first test for this optimization method was to measure the CPU time of a shear simulation for grains of different number of sides. We start with triangles and end with heptagons. the results are shown in Fig. \ref{fig:geogranu}. Initially with an $\alpha=0.9$ the Verlet list is not frequently updated and hence the optimization is for all purposes turned off. The force is then the predominant load on the CPU time and its calculation grows like a quadratic law with the numbers of sides.

This change with the introduction of a smaller $\alpha$ value. The optimization reduce the force load on the total calculation time. The power law dependence observed in the previous case is changed (exponent is equal to 0.3533), and there is no apparent difference between the time expended in the calculation for hexagons and for heptagons. With the optimization the time no longer grows in a quadratic form.

\begin{figure}[H]
 \begin{center}
 \includegraphics[bb = 0 0 1140 1614,scale=0.4,trim=0cm 0cm 15cm 35cm,clip]{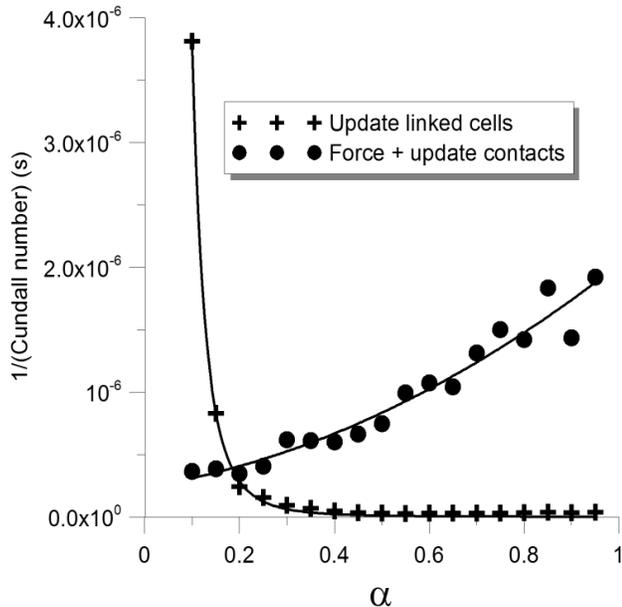}
 \caption{ The distinct times taken for a shear band simulation with disks elements with the profiling utility of the gcc compiler versus the $\alpha$ parameter}
 \label{disksgranu}
 \end{center}
 \end{figure}

The next test was to measure the specific time that the simulation assign in both calculating forces and updating the Verlet list. The results for disks are show in Fig. \ref{disksgranu}.

The total CPU time is mainly split in force calculation and updating the Verlet lists. The time spent in each one of these features depends on $\alpha$. Now for a small $\alpha$ the Verlet list are more frequently updated, and actually for $\alpha \rightarrow 0$ the updating frequency will diverge to $\infty$ also. In contrast when $\alpha$ is large, the simulation no longer updates the Verlet list and therefore the time spent in this action goes to zero. Hence an invert proportional relation should be adequate to model this part of the total time. Now the time spent in the force calculation grows with $\alpha$. We propose a quadratic law since the number of grains in a Verlet distance to the particle will be proportional to its area.

\begin{figure}[h]
 \begin{center}
 \includegraphics[bb = 0 0 1140 1614,scale=0.4,trim=0cm 0cm 15cm 35cm,clip]{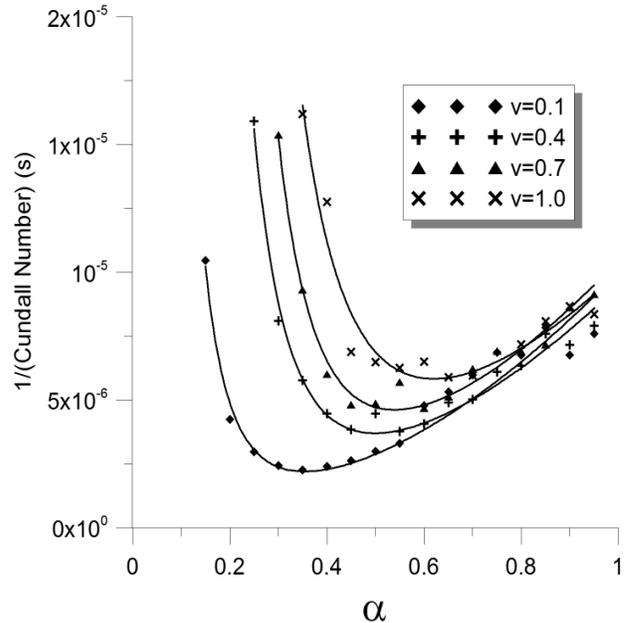}
 \caption{ Obtained cpu time for the shear band test with different shear rates. The fitted curves are obtained with the model proposed on Eq. \ref{fm}}
\label{cpuvsv}
 \end{center}
 \end{figure}

With this idea me made the final test to our method: The construction of a shear band simulation (see Fig. \ref{fig:shear}) in which the load is moving to the right side with a constant velocity. We check the CPU time vs the velocity of the load for a range of $\alpha$ values going from 0.1 to 0.9 and with three different load velocities (0.1, 0.4,0.7 and 1.). The results are shown in Fig. \ref{cpuvsv}.

The results where fitted with this analytical model relating the CPU time with the verlet length $\alpha$:
\begin{equation}\label{fm}
T_{CPU}=T_{uc}+T_{f}=\frac{A}{\alpha ^\beta}+B \alpha ^2
\end{equation}

Where $T_{uc}$ is the time spent on updating the verlet list, $T_{f}$ is the time for calculating the forces, $A$, $B$ and $\beta$ are de fitting parameters. As we can see the optimum value value for $\alpha$ change with the shear rate as well as the total CPU time for a given $\alpha$. For the lowest shear rate we have the best $\alpha$ value, which means that the Verlet list method accomplishes a greater optimization if the particles move slowly.

\section{Concluding Remarks}
\label{conclusions}

The main motivation of this paper was to show that the concepts of computational geometry (Voronoi
diagrams, minkowski operator and distance formulas) are useful to model systems composed with
complex shaped particles. These concepts are applied to simulate systems of spheropolygons interacting via
dissipative forces. We guarantee energy balance, which were not available in previous models.
The classical concept of Verlet list was extended for spheropolygons. Depending on the
characteristic velocity of the system there is an optimal value of the Verlet distance that
minimize CPU time.

The most promising aspect of this work is its natural extension to 3D. There are several available
subroutines to perform Voronoi Tesselation in 3D, which will allow to generate the Voronoi-Minkowski
diagrams once the subroutine of erosion is implemented. The analytical calculation of the mass
properties in 3D spheropolyhedra can be obtained by decomposing them in polyhedra and spherical
segments. The contact force between spheropolyhedra can be calculated as the sum of all vertex-face
and edge-edge contacts. The equation of motion for the rotational degrees of freedom are not too
trivial like in 2D, but the quaternion formalism \cite{wang2006ips} can be used to a correct description of all the
degrees of freedom.

\acknowledgments
We thank Y.C. Wang for   discussions.  This work is supported by UQ ECR project
and the Australian Research Council (project number DP0772409 )

\end{document}